\documentclass[a4paper,11pt]{article}
\usepackage{pos}

\title{Forward dijet and vector meson production at the LHC}
\author*[a]{Christophe Royon}
\author[a,b]{Jani Penttala}

\affiliation[a]{Department of Physics and Astronomy,\\
 The University of Kansas, Lawrence, KS 66045, USA}

\affiliation[b]{Department of Physics and Astronomy,\\
University of California, Los Angeles, CA 90095, USA}

\emailAdd{christophe.royon@ku.edu}
\emailAdd{janipenttala@physics.ucla.edu}

\abstract{We first describe the predictions concerning jet production in the very forward region, especially in the FOCAL kinematic region of the ALICE collaboration at CERN as a test of saturation. We also compare vector meson production cross sections at the LHC in Pb Pb and p Pb collisions with the Balitsky Fadin Kuraev Lipatov and the Balitsky Kovchegov evolution equations and data favor the observation of saturation at the LHC.}

\FullConference{42nd International Conference on High Energy Physics (ICHEP2024)\\
18-24 July 2024\\
Prague, Czech Republic\\}

\begin{document}
\maketitle

In this short report, we will describe two observables that are sensitive to Balitsky Fadin Kuraev Lipatov (BFKL)~\cite{bfkl} dynamics and to saturation effects in heavy ion collisions at the LHC, namely very forward jet and vector meson productions.

\section{Very forward jet production in $pA$ collisions at the LHC}

The first observable sensitive to saturation effects that we are going to discuss is the possible measurement of very forward jet production in $p Pb$ collisions at the LHC. In order to be sensitive to saturation effects, one needs to probe the gluon density in the $Pb$ ion at small $x$. in $pPb$ collisions, when both jets are emitted in the very forward regions, we can reach values of $x$ close to 10$^{-5}$ on the $Pb$ side. In the following, we will thus consider jet measurements in the forward region of the CMS and ATLAS detectors called ``forward" CMS kinematics~\cite{forwardcms} ($3.5 < y_{jet} <4.5$  and $p_T^{jet}$ between 10 and 20 GeV,  20 and 40 GeV or 40 and 80 GeV) and ``CASTOR/FOCAL"~\cite{focal} kinematics ($5.2 < y_{jet} <6.6$ and $p_T^{jet}$ between 5 and 10 GeV, or 10 and 20 GeV)~\cite{ourpapb}.

In order to predict the forward jet cross section, our model starts by factorizing the photon into a $q \bar{q}$ pair from the $q \bar{q}$ scattering off a dense nuclear target, such as $Pb$. We use the dipole amplitude fitted to HERA data from the AAMQS (A non-linear QCD analysis of new HERA data at small-$x$)~\cite{aamqs} parametrization.  The Balitsky Kovchegov (BK)~\cite{bk} equation is used to evolve the dipole density at small $x$. Unfortunately, the AAMQS parametrization does not contain any $b$ impact parameter dependence, which is needed to get precise predictions. We thus use the IPSat model~\cite{ipsat} to obtain the impact parameter dependence. However, the IPSat model does not contain any BK evolution and our model is a ``mixture" of the AAMQS and IPSat models~\cite{ourpapb}. The additional original  aspect of our model is that we consider the sum of each proton and neutron gaussian thicknesses as the $Pb$ thickness
\begin{eqnarray}
T(b) = \Sigma_{i=1}^A T_{p/n} (b_i-b) \nonumber
\end{eqnarray}
where the nucleon impact parameters ($b_i$'s) are generated stochastically.

\begin{figure}[t!]
    \centering
    \includegraphics[width=0.72\textwidth]{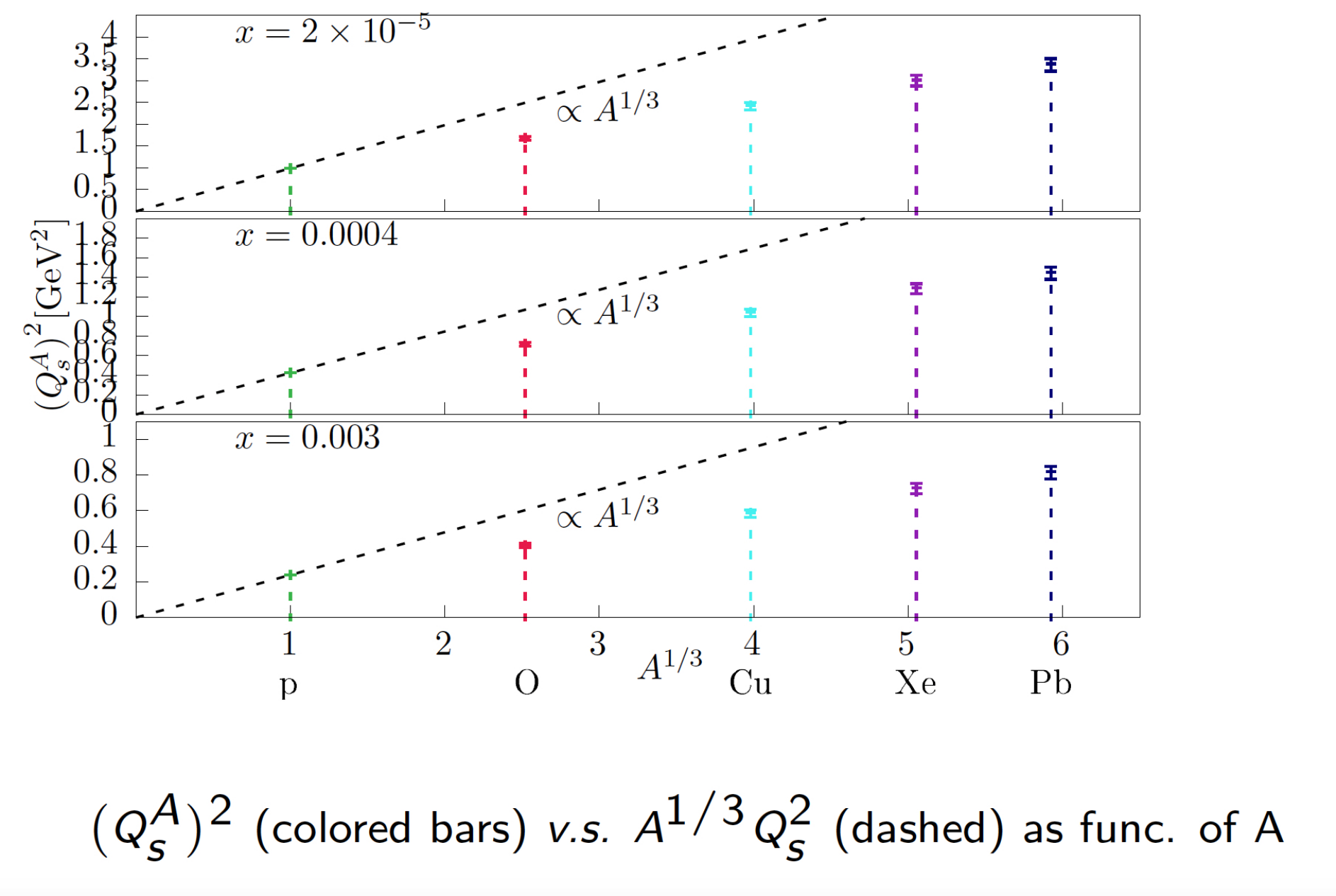}
    \caption{Saturation scale $(Q^A_S)^2$ (colored bars) vs naive expectations $(A^{1/3} Q_S^2$)
        }
    \label{satscale}
\end{figure}

After checking that the $F_2$ structure function description as measured at HERA is similar if we use the $b$-independent AAMQS parametrization or our own parametrization after integration on the $b$-parameter, we can predict the forward jet cross sections and the values of the saturation scales. We first compare the saturation scales given by our model (using the $b$-dependence) and the naive one with the usual $A^{1/3}$ dependence in Fig.~\ref{satscale} for $p$, $O$, $Cu$, $Xe$ and $Pb$ for three $x$ values ($x=2~10^{-5}$, 0.0004 and 0.003) and we obtain lower saturation scales than using the naive $A^{1/3}$ dependence.

\begin{figure}[t!]
    \centering
    \includegraphics[width=0.49\textwidth]{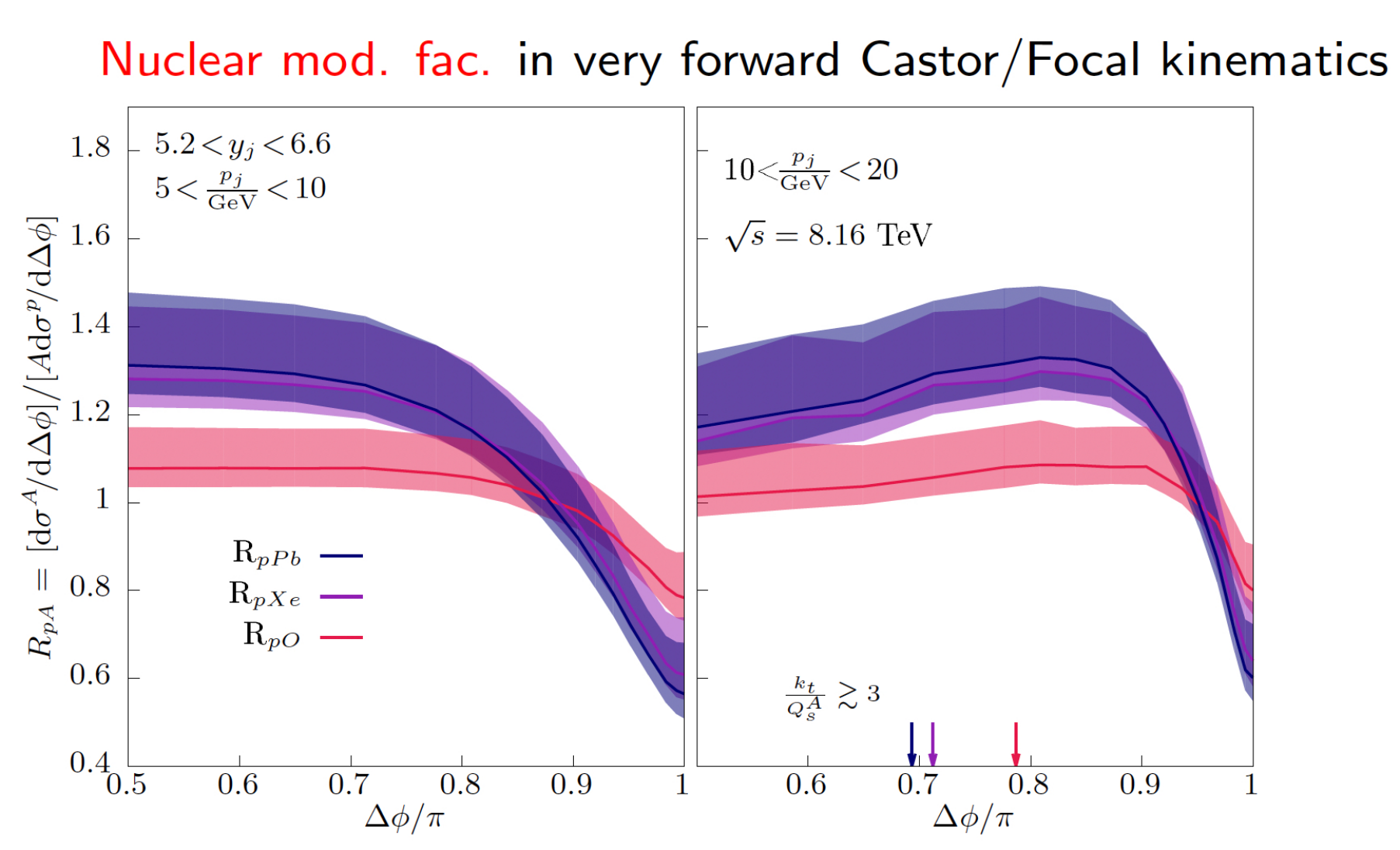}
    \includegraphics[width=0.49\textwidth]{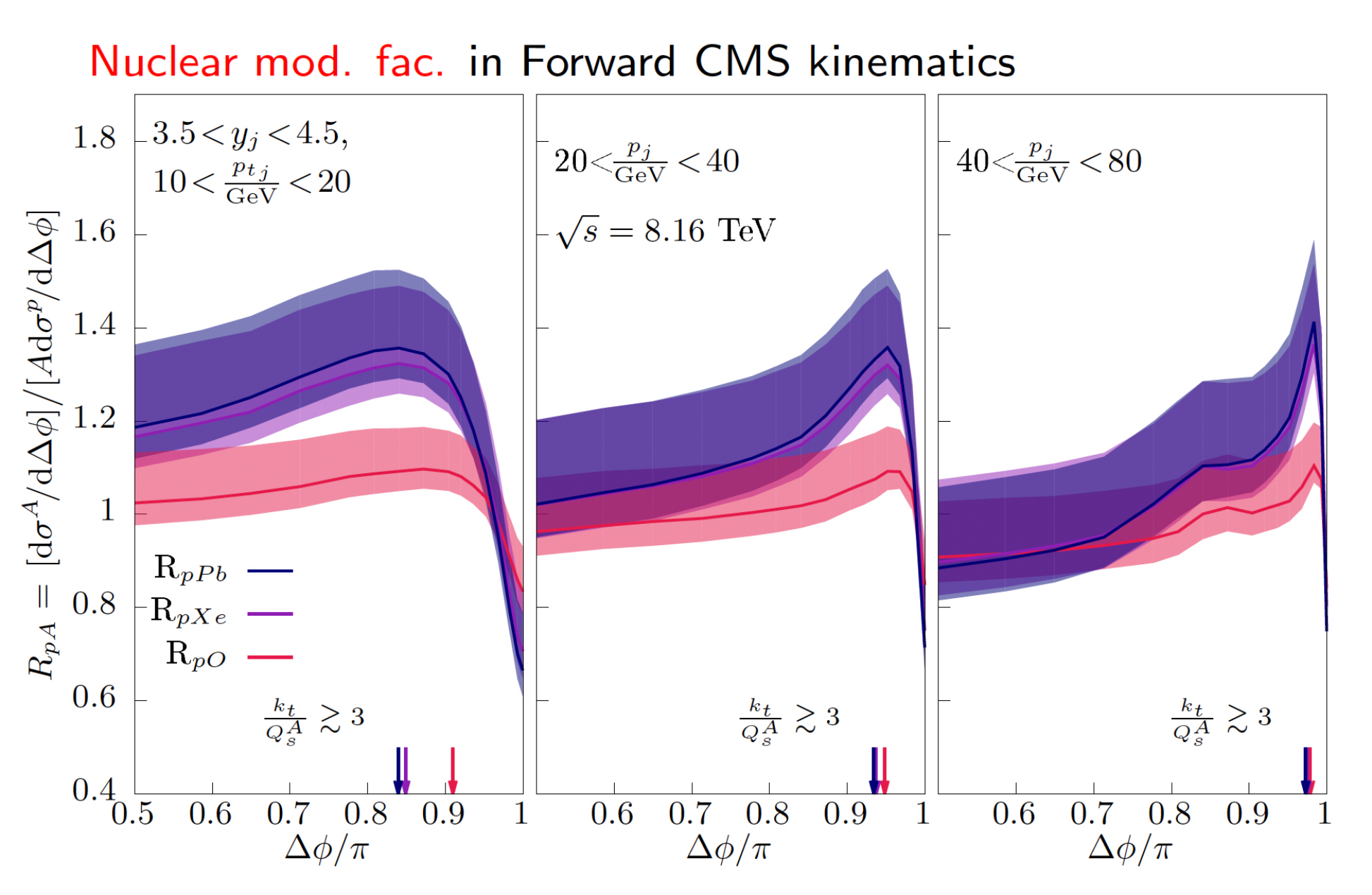}
    \caption{Nuclear modification factors for two detector configurations (CASTOR/FOCAL and FORWARD-CMS) for different $p_T$ intervals.
        }
    \label{rpa}
\end{figure}

\begin{figure}[t!]
    \centering
    \includegraphics[width=0.49\textwidth]{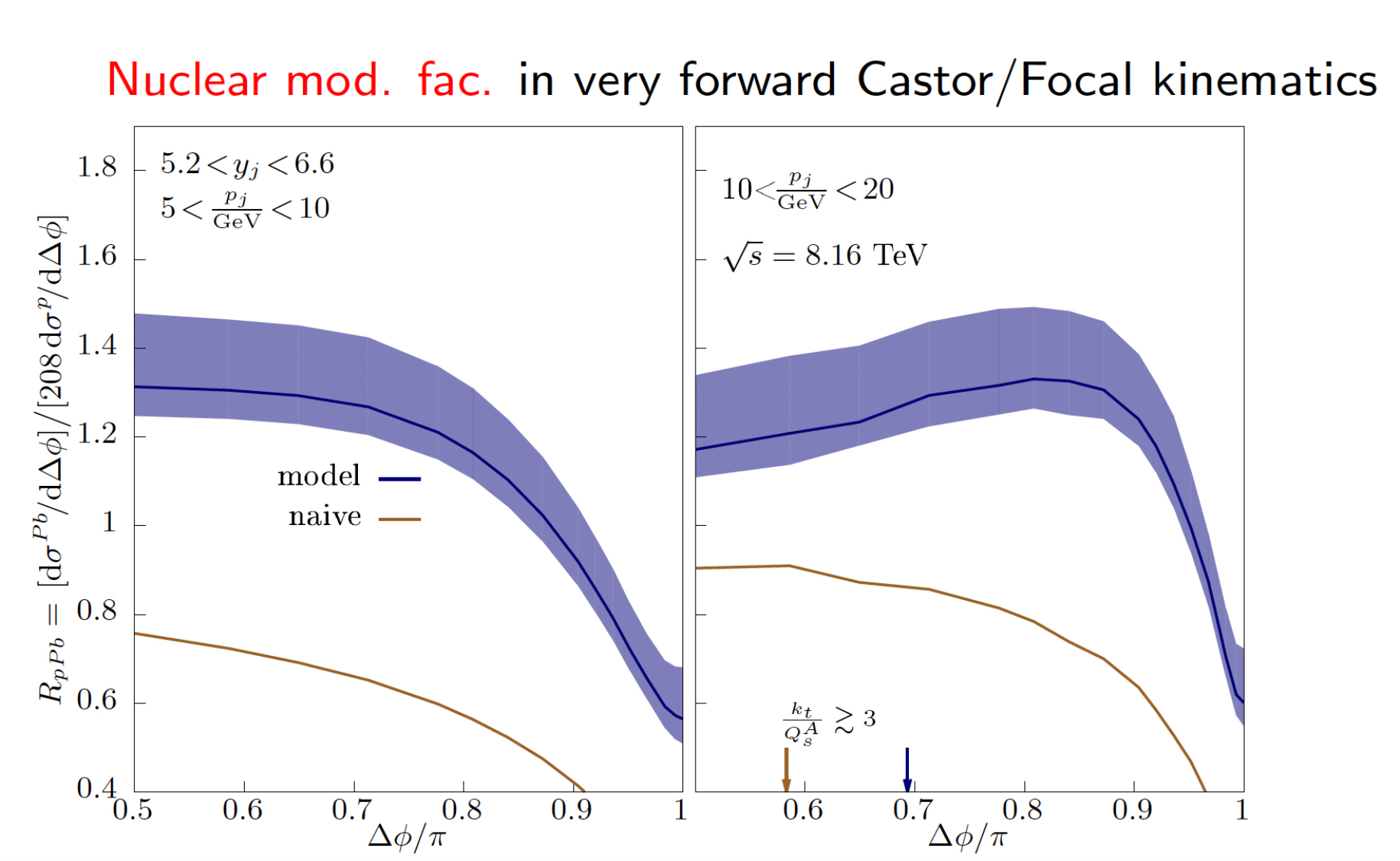}
    \includegraphics[width=0.49\textwidth]{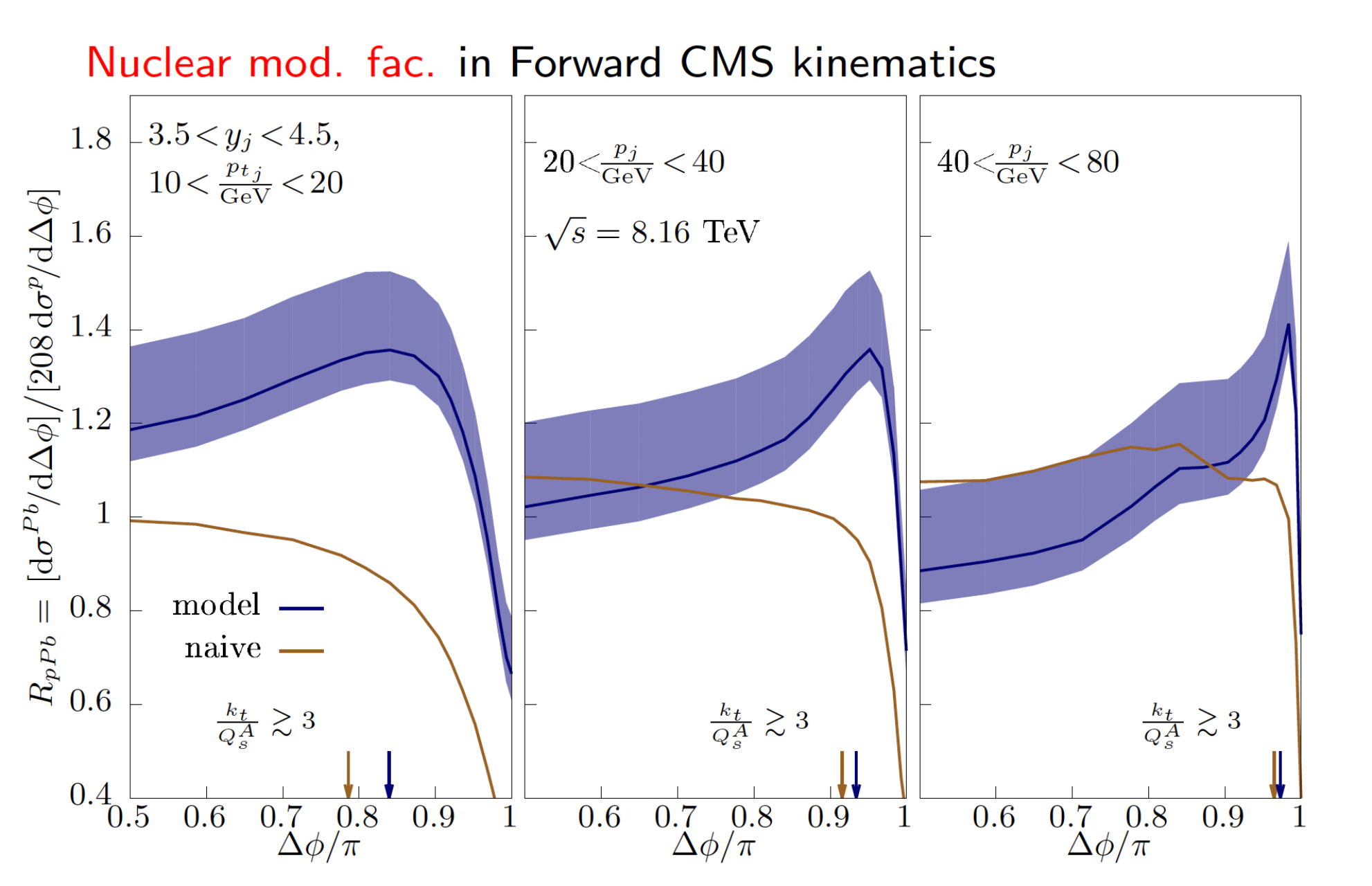}
    \caption{Nuclear modification factors for two detector configurations (CASTOR/FOCAL and FORWARD-CMS) for different $p_T$ intervals compared with the naive expectations.
        }
    \label{rpab}
\end{figure}

The nuclear modification factors for the two detector configurations, namely the ``forward CMS" and ``CASTOR FOCAL" kinematics are shown in Figs~\ref{rpa} and \ref{rpab} as a function of the azimuthal angle for $Pb$, $Xe$ and $O$ between the two forward jets. $Pb$ and $Xe$ lead to similar decorrelations. As expected, the azimuthal decorrelation is more pronounced for higher $y$ and lower $p_T$.  We also observe a large difference between our model and the naive one. It will be quite relevant to test our model using especially the FOCAL forward detector to be installed in ALICE.

\section{Vector meson production in $p Pb$ and $Pb Pb$ collisions at the LHC}

\subsection{What do we need to see saturation at the LHC?}

In order to see some evidence of saturation at the LHC in the existing data, one needs to probe a dense object, so a heavy ion such as $Pb$. Exclusive productions of vector mesons ($J/\Psi$, $\Upsilon$) or of $c \bar{c}$ for instance in ultra-peripheral collisions (UPC) are dominated by $\gamma$-Pomeron interactions where we probe the gluon distributions in the $Pb$ ions.  These are ideal probes for low-$x$ physics where one  can reach low $x$ values of 10$^{-5}$ or smaller when particles are produced at high rapidities.
In addition, in order to look for gluon saturation effects, one needs to probe a low scale (to be below the saturation scale $Q_S$) while still being in the perturbative region, and this is why $c$ or $b$ productions, where one can go to very low $p_T$, or $J/\Psi$ (low mass vector mesons) are ideal probes.

Our idea is thus to compute exclusive vector meson production in $\gamma p$ (HERA, EIC and pPb LHC)
and $\gamma Pb$ (EIC and Pb Pb LHC) interactions where we probe the gluon density in $p$ or $Pb$.
Saturation effects are expected to happen in $Pb Pb$ collisions when one probes the gluon density at high energies $W$, so small $x$, but not in $p Pb$ collisions where one probes the gluon density in the proton (the $Pb$ ion emitting the quasi-real photon).

\begin{figure}[t!]
    \centering
    \includegraphics[width=0.52\textwidth]{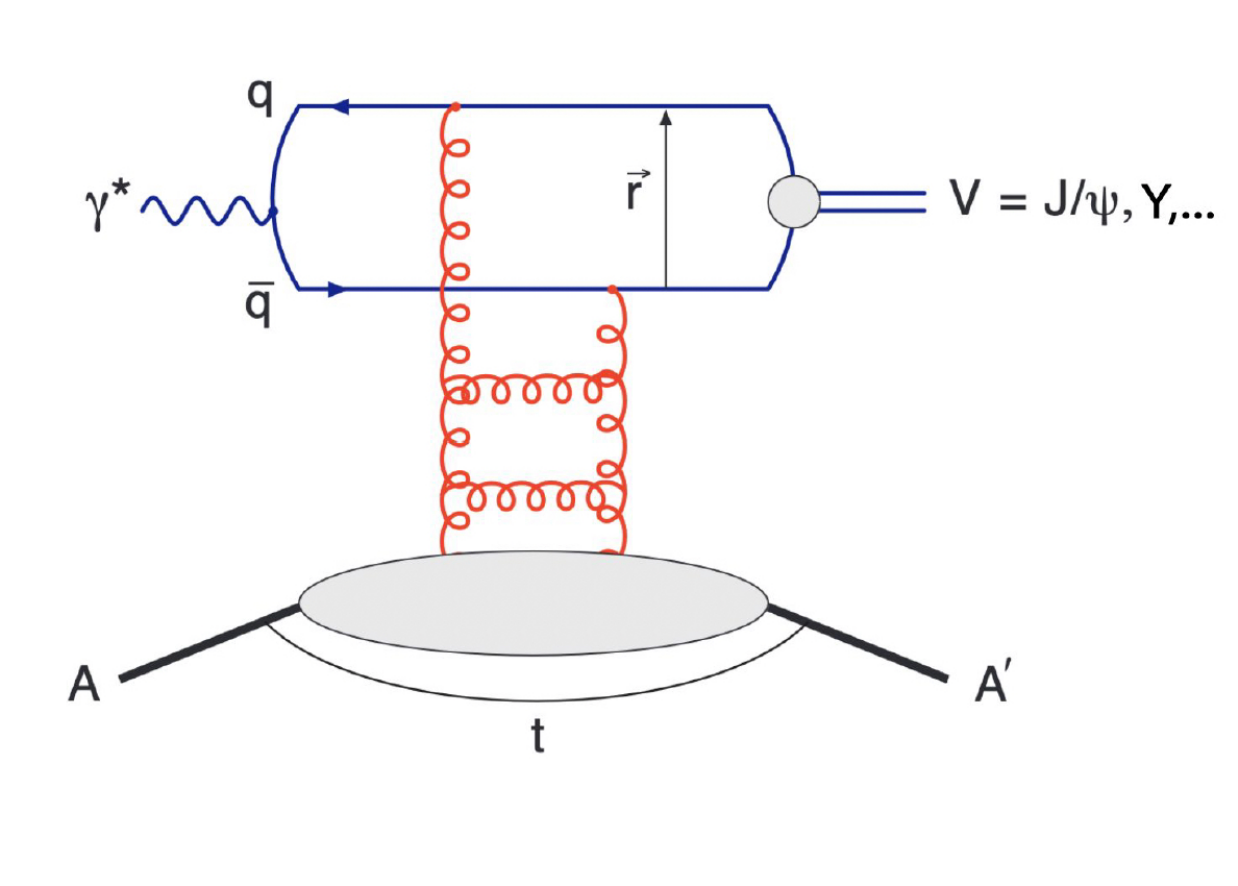}
    \caption{Exclusive vector meson production in $Pb Pb$ collisions.
        }
    \label{vectormeson}
\end{figure}

In the following, we will compute the vector meson production cross section by factorizing the $\gamma \rightarrow q \bar{q}$ part from the dipole density in protons or $Pb$ as illustrated in Fig.~\ref{vectormeson}. We use the linear BFKL and the non-linear BK evolution equations including saturation effects to evolve the dipole densities. 
In addition, we take into account $b$ impact parameter dependence in the dipole amplitude including a gaussian dependence of the thickness function for protons and the Wood-Saxon formalism for $Pb$~\cite{salazar}. Taking into account the $b$-dependence of the dipole amplitude is crucial. 

\subsection{$J/\Psi$ UPC production at the LHC}

\begin{figure}[t!]
    \centering
    \includegraphics[width=0.32\textwidth]{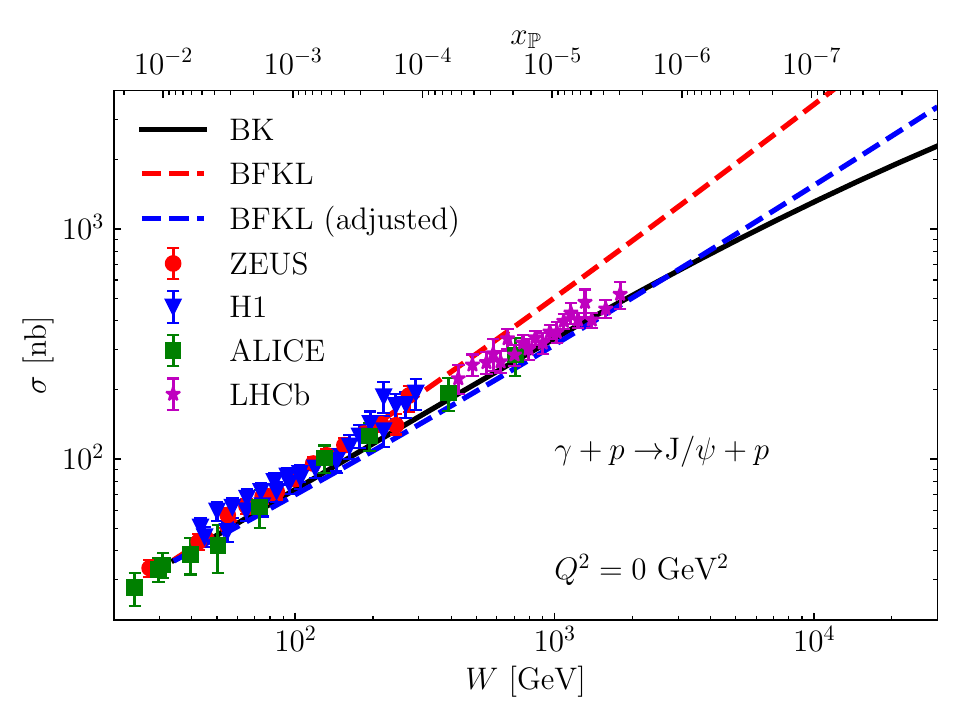}
    \includegraphics[width=0.32\textwidth]{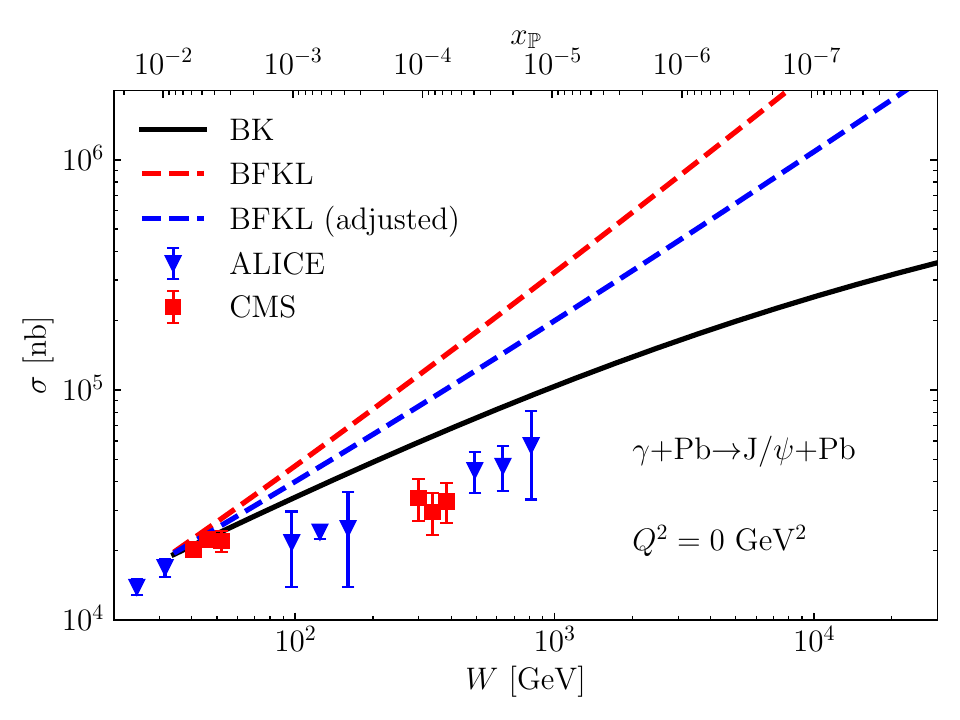}
    \includegraphics[width=0.32\textwidth]{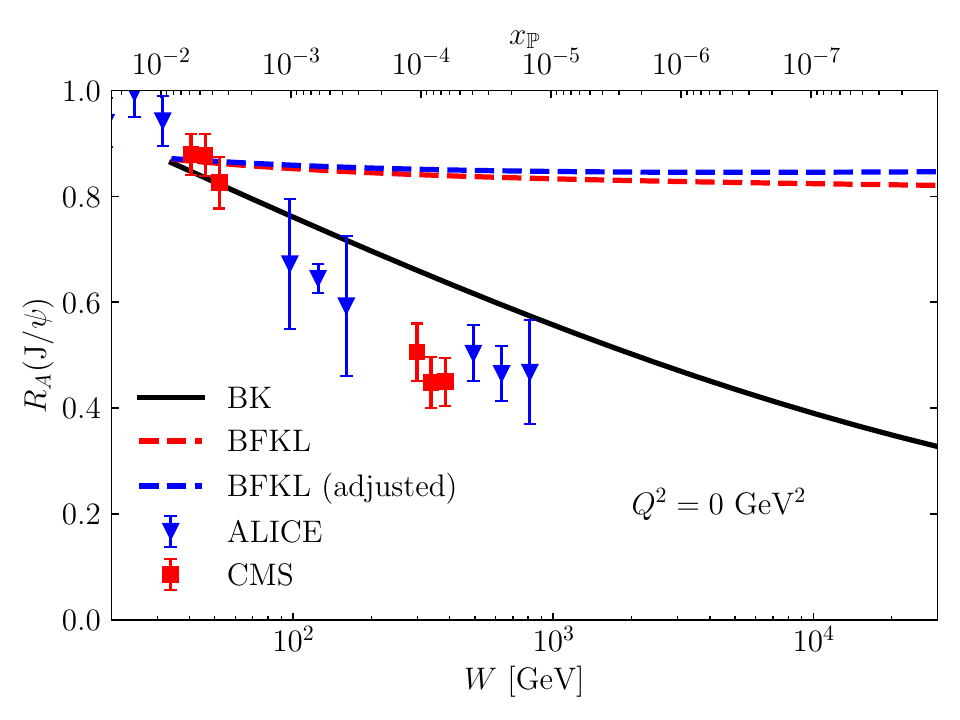}
    \caption{Exclusive $J \Psi$ production as a function of the center-of-mass energy $W$. 
    Left: Proton target, 
    Center: Lead target,
  Right: Nuclear suppression factor,
        }
    \label{jpsi}
\end{figure}

The results for the $\gamma p$ and $\gamma Pb$ cross section predictions for $J/\Psi$ production are shown in Fig.~\ref{jpsi}, left and center. They are compared with the data from the H1~\cite{h1}, ZEUS~\cite{zeus} collaborations at HERA at lower energies and from the CMS~\cite{cms}, ALICE~\cite{alice}, LHCb~\cite{lhcb} collaborations at the LHC in $p Pb$ and $Pb Pb$ collisions. 

The $J/\Psi$ production in $pPb$ shows small differences between BK and BFKL resummations as expected (the ``adjusted"  BFKL predictions correspond to a fit of the $\alpha_S$ value to the vector meson data).  The same parameters are kept for the BK evolution. Large differences are observed between the BK and BFKL predictions in PbPb collisions and data favor the BK evolution. Data thus favor the presence of saturation in Pb Pb collisions at high $W$ at the LHC. It is worth noting that the main differences between the BFKL and BK approaches rely on the linearity of the $W$ dependence in the logarithmic scale for the BFKL evolution whereas the slope changes for BK. This property is quite stable even if one includes higher order corrections whereas the absolute normalization might differ. It is also worth noticing that saturation effects seem to be more important in data than predicted by BK (it might be due to higher order effects).
In Fig.~\ref{jpsi}, right, the large nuclear suppression factor for $J/\Psi$ in $PbPb$ collisions is also displayed.

\subsection{$\Upsilon$ UPC production at the LHC}

\begin{figure}[t!]
    \centering
    \includegraphics[width=0.32\textwidth]{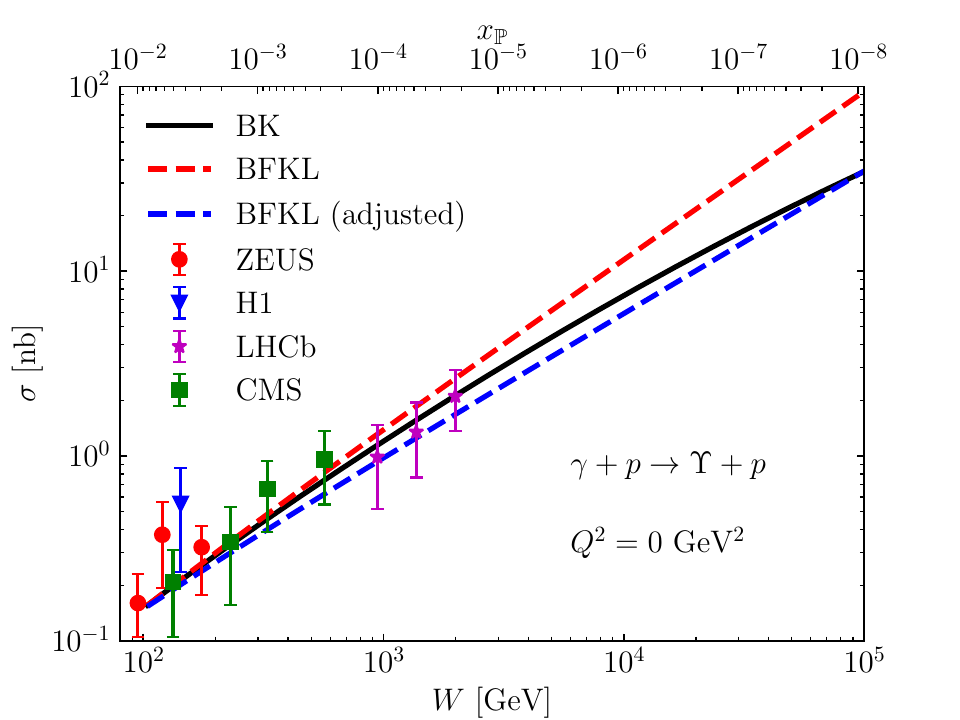}
    \includegraphics[width=0.32\textwidth]{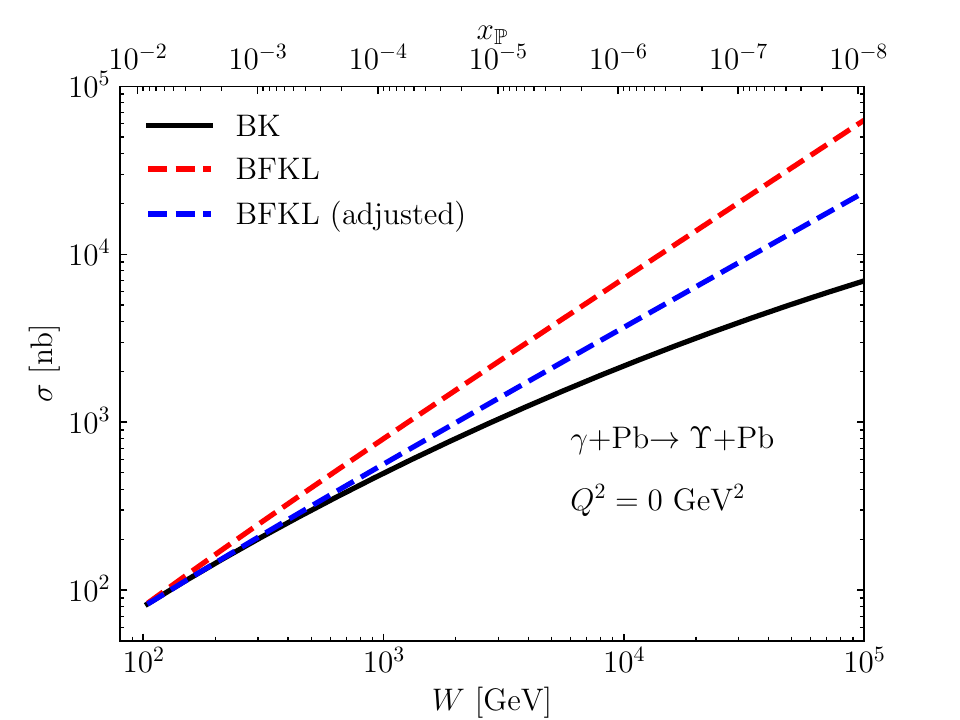}
    \includegraphics[width=0.32\textwidth]{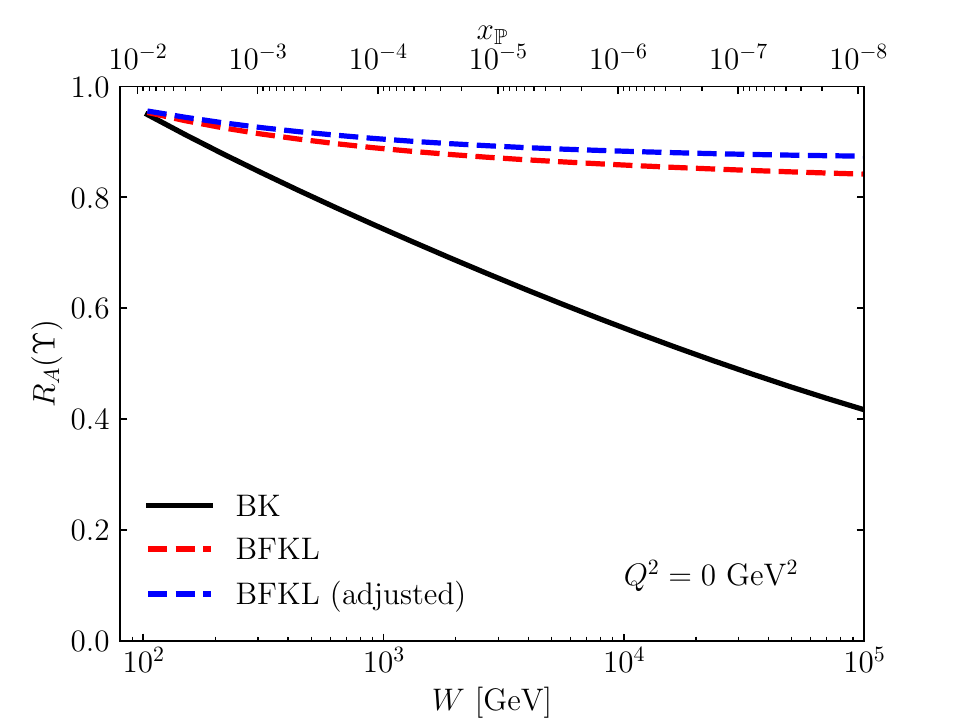}
    \caption{Exclusive $\Upsilon$ production as a function of the center-of-mass energy $W$. 
    Left: Proton target,
    Center: Lead target.
    Right: Nuclear suppression factor.}
    \label{upsilon}
\end{figure}

The results for the $\gamma p$ and $\gamma Pb$ cross section predictions for $\Upsilon$ production are shown in Fig.~\ref{upsilon}, left and center.  BFKL and BK approaches lead to similar cross sections for $\gamma p$ interactions and we observe smaller differences between BFKL and BK in $\gamma Pb$ interactions than for $J/\Psi$ because of the higher mass of the $\Upsilon$ meson. A precise measurement of $\Upsilon$ production in UPC $Pb Pb$ collisions is thus of high interest since it corresponds to the transition region between the saturation and dilute regimes. In Fig.~\ref{upsilon}, right, the nuclear suppression factor for $\Upsilon$ in $PbPb$ collisions is displayed.

\section{Conclusion}

In this short report, we first predicted the forward jet azimuthal cross sections for different heavy ions using a $b$-dependent approach and considering the sum of each proton and neutron gaussian thicknesses as the heavy ion thickness. It leads to lower decorrelation between jets and to lower saturation scale than the simple  $A^{1/3}$ dependence. The second part of our work is the comparison between $J/\Psi$ and $\Upsilon$ UPC measurements in $ep$ measurements at HERA and $p Pb$ and $Pb Pb$ measurements at the LHC and BFKL and BK predictions. Data show a clear preference of saturation effects at high energy for $Pb Pb$ collisions.

%\section*{Acknowledgments}
%The results on vector mesons come from a fruitful collaboration with Jani Penttala.

\end{document}